\documentclass[3p,twocolumn]{elsarticle}

\usepackage{lineno,hyperref}
\usepackage{amsmath}
\usepackage{booktabs}
\usepackage{amssymb}
\usepackage{multirow}
\usepackage{color}
\usepackage{soul}

\newcommand{\nuc}[2]{$^{#1}$#2}
\newcommand{\diff}{\text{d}}

\newcommand{\asymmerror}[3]{$#1^{+#2}_{-#3}$}
\usepackage{lineno,hyperref}
\modulolinenumbers[5]

\journal{Physics Letters B}

\begin{document}

\begin{frontmatter}

\title{Compressional-mode resonances in the molybdenum isotopes: Emergence of softness in open-shell nuclei near A=90}

\author[ND]{K.B. Howard}
\author[ND]{U. Garg}
\author[CYRIC]{M. Itoh}
\author[Konan]{H. Akimune}
\author[RCNP]{M. Fujiwara}
\author[Kyoto,RCNP]{T. Furuno}
\author[ND,BARC]{Y.K. Gupta}
\author[KVI]{M.N. Harakeh}
\author[Kyoto]{K. Inaba}
\author[CYRIC]{Y. Ishibashi}
\author[CYRIC]{K. Karasudani}
\author[Kyoto,Toyonaka]{T. Kawabata}
\author[RCNP]{A. Kohda}
\author[CYRIC]{Y. Matsuda}
\author[Kyoto]{M. Murata}
\author[RCNP]{S. Nakamura}
\author[CYRIC]{J. Okamoto}
\author[CNS]{S. Ota}
\author[FSU]{J. Piekarewicz}
\author[Kyoto]{A. Sakaue}
\author[ND,Ankara]{M. \c{S}enyi\u{g}it}
\author[Kyoto]{M. Tsumura}
\author[ND]{Y. Yang}

\address[ND]{Department of Physics, University of Notre Dame, Notre Dame, Indiana 46556, USA}
\address[CYRIC]{Cyclotron and Radioisotope Center, Tohoku University, Sendai 980-8578, Japan}
\address[Konan]{Department of Physics, Konan University, Kobe 658-8501, Japan}
\address[RCNP]{Research Center for Nuclear Physics, Osaka University, Osaka 567-0047, Japan}
\address[Kyoto]{Department of Physics, Kyoto University, Kyoto 606-8502, Japan}
\address[BARC]{Nuclear Physics Division, Bhabha Atomic Research Centre, Mumbai 400085, India}
\address[KVI]{KVI-CART, University of Groningen, 9747 AA Groningen, The Netherlands}
\address[Toyonaka]{Department of Physics, Osaka University, Toyonaka, Osaka 560-0043, Japan}
\address[CNS]{Center for Nuclear Study, University of Tokyo, RIKEN Campus, Wako, Saitama 351-0198, Japan}
\address[FSU]{Department of Physics, Florida State University, Tallahassee, Florida 32306, USA}
\address[Ankara]{Department of Physics, Faculty of Science, Ankara University, TR-06100, Tando\^{g}an, Ankara, Turkey}

\begin{abstract}
  ``\emph{Why are the tin isotopes soft?}'' has remained, for the past decade, an open problem in nuclear structure physics: models which reproduce the isoscalar giant monopole resonance (ISGMR) in the ``doubly-closed shell''
nuclei, \nuc{90}{Zr} and \nuc{208}{Pb}, overestimate the ISGMR energies of the open-shell tin and cadmium nuclei, by as much as 1 MeV. In an effort to shed some light onto this problem, we present results of detailed studies of the ISGMR in the molybdenum nuclei, with the goal of elucidating where--and how--the softness manifests itself between \nuc{90}{Zr} and the cadmium and tin isotopes. The experiment was conducted using the \nuc{94,96,98,100}{Mo}($\alpha,\alpha^\prime$) reaction at $E_\alpha = 386$ MeV. A comparison of the results with relativistic, self-consistent Random-Phase Approximation calculations indicates that the ISGMR response begins to show softness in the molybdenum isotopes beginning with $A=92$.
\end{abstract}

\begin{keyword}
Collectivity, giant resonance, nuclear incompressibility, softness, equation of state
\end{keyword}

\end{frontmatter}

The compressional-mode isoscalar giant monopole resonance (ISGMR) has long been regarded as an optimal experimental probe for constraining the equation of state (EoS) of nuclear matter close to saturation density~\cite{harakeh_book,piekarewicz_correlating_GMR_to_incompressibility,cao_sagawa_colo,garg_colo_review}. In particular, the nuclear incompressibility, $K_\infty$, has been shown to be strongly correlated with properties of ISGMR so that measurements of the excitation energies of the ISGMR in finite nuclei can be used to infer directly the value of $K_\infty$~\cite{cao_sagawa_colo,garg_colo_review,blaizot_nuclear_compressibilities,blaizot,colo_2004a}.
The mechanism through which this occurs is detailed in Refs.~\cite{harakeh_book,cao_sagawa_colo,garg_colo_review,blaizot} and is predicated on the assumption that close to $100\%$ of the energy-weighted sum rule (EWSR) is exhausted within a single collective peak in the experimentally extracted ISGMR strength distribution~\cite{harakeh_book}. One builds a class of
interactions --- calibrated to the ground-state properties of finite nuclei --- which span a wide range of possible $K_\infty$ values and, correspondingly, makes predictions for the ISGMR energies for a given nucleus. Within this framework, one takes the experimentally extracted ISGMR and finds which value of $K_\infty$ best reproduces the measured excitation energies. This procedure is generally understood to be insensitive to the choice of the nucleus made in determining the true value for $K_\infty$ for bulk nuclear matter~\cite{harakeh_book}.
With this  prescription applied to the ``doubly-closed'' nuclei, \nuc{208}{Pb} and \nuc{90}{Zr}, a value of $K_\infty = 240 \pm 20$ MeV has been established using a myriad of relativistic and non-relativistic interactions~\cite{colo_2004a,todd-rutel_piekarewicz_relativistic_interactions_and_neutron_stars,shlomo_knm_240,piekarewicz_symmetry_energy,Colo2014a}.

It was, therefore, very puzzling when a series of ISGMR measurements in the tin isotopes~\cite{Li_PRL,Li_PRC} appeared to be inconsistent with this adopted value for $K_\infty$. It was found that in the \nuc{112-124}{Sn} isotopic chain, the predicted ISGMR energies were systematically overestimated by several hundred keV; the interpretation of this phenomenon is, quite simply, that the tin isotopes appear to be ``soft'' in relation to \nuc{208}{Pb} and \nuc{90}{Zr} insofar as the ISGMR energies are concerned~\cite{piek_fluffy_physRev,garg_fluffy_nucA}. Indeed, the values of $K_\infty$ which would be extracted using the tin isotopes as the benchmark, would be well below the suggested range of
$K_\infty = 240 \pm 20$ MeV~\cite{piek_fluffy_physRev}.
The cadmium nuclei were also found to exhibit the same behavior~\cite{patel_cd}.

A number of possible solutions were proposed to explain this incomprehensible softness, such as the notion of mutually-enhanced magicity (MEM) in doubly-closed shell nuclei~\cite{khan_MEM}, as well as contributions due to superfluid pairing interactions~\cite{junli_pairing,khan_pairing,tselyaev_Sn_quasiparticle}. The MEM effect was refuted by experimental observations by Patel \emph{et al.}~\cite{patel_MEM}, and the exact effects of pairing on the ISGMR are still somewhat uncertain~\cite{khan_pairing}, but nonetheless have been determined to be insufficient for accounting for this softening.

This puzzle is deemed a fundamental open problem in nuclear structure physics and remains unresolved to this day~\cite{cao_sagawa_colo,Li_PRL,Li_PRC,piek_fluffy_physRev,garg_fluffy_nucA,patel_cd,tselyaev_Sn_quasiparticle,piek_physG,vesely_phys_rev_c}.
This question can naturally be extended to the molybdenum ($Z\!=\!42$) isotopic chain: put simply,
if the tin and cadmium isotopes ($Z\!=\!50$ and $48$, respectively) are soft relative to
\nuc{90}{Zr} ($Z\!=\!40$), where the ISGMR response in zirconium is consistent with that in \nuc{208}{Pb},
then what changes occur  between zirconium and cadmium (and tin) and where does this softening emerge?
In this Letter, we report on a systematic study of the
%giant resonances in the
molybdenum isotopes with the goal of elucidating whether the excitation energies of the ISGMR soften in moving away from
the ``doubly-magic'' nucleus \nuc{90}{Zr}. The results provide clear evidence that this softening develops as nucleons are added to \nuc{90}{Zr}, and that \nuc{94,96}{Mo} are similar to the cadmium and tin isotopes as far as an extraction of $K_\infty$ from the corresponding ISGMR energies is concerned.

The measurements were carried out at the Research Center for Nuclear Physics (RCNP) at Osaka University. A high-quality ``halo-free'' beam of $386$-MeV $\alpha$-particles impinged onto enriched (90--95\% isotopic purity), self-supporting \nuc{94,96,98,100}{Mo} targets with areal densities $\simeq$\,1--5\,mg/cm$^2$. Inelastically-scattered $\alpha$-particles were momentum-analyzed in the high-resolution spectrometer Grand Raiden, and transported to the focal-plane detection system in vertical-focusing mode~\cite{tamii_grand_raiden,Fujiwara_Grand_Raiden,von-neumann-cosel_tamii_RCNP}. The detection system consisted of a pair of multiwire drift chambers with both vertical and horizontal position sensitivity, as well as a pair of plastic scintillators that provided the particle identification. A detailed description of the data reduction process is provided in Refs.~\cite{KBH_EPJA,KBH_thesis}. Here, we will only reiterate the major benefit provided by the ion-optics of Grand Raiden in vertical focusing mode: this arrangement renders it possible to subtract out the contributions from the instrumental background events, which underlie the true inelastic scattering spectra, based on their detected vertical positions. This feature is unique to measurements of the ISGMR at RCNP, as otherwise one is forced to account for the instrumental background and physical continuum in a phenomenological way, often with ill effect on the extracted ISGMR strength~\cite{KBH_EPJA,TAMU_40Ca,TAMU_48Ca,youngblood_A90_unexpected,TAMU_44Ca,KBH_calcium_PLB}.

\begin{figure}[h!]
  \includegraphics[width=\linewidth]{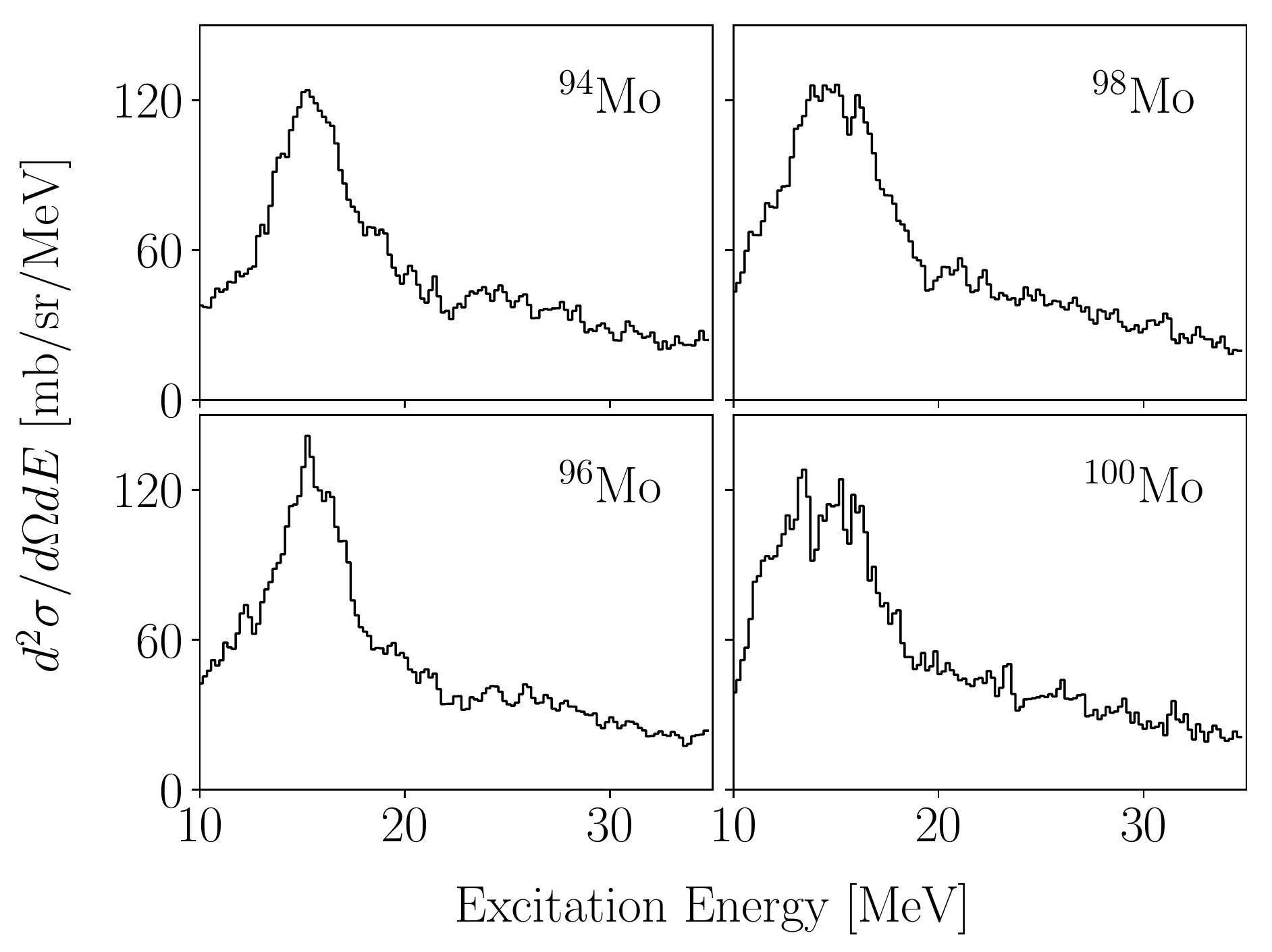}
  \caption{Measured double-differential cross-section spectra from \nuc{94,96,98,100}{Mo}($\alpha,\alpha^\prime$) at an average scattering angle of $0.7^\circ$ (\emph{i.e.} averaging over the opening angle of the spectrometer set at $0^\circ$), after particle identification and subtraction of the instrumental background.}
  \label{molly_spec}
\end{figure}

\begin{figure*}[t!]
  \includegraphics[width=\linewidth]{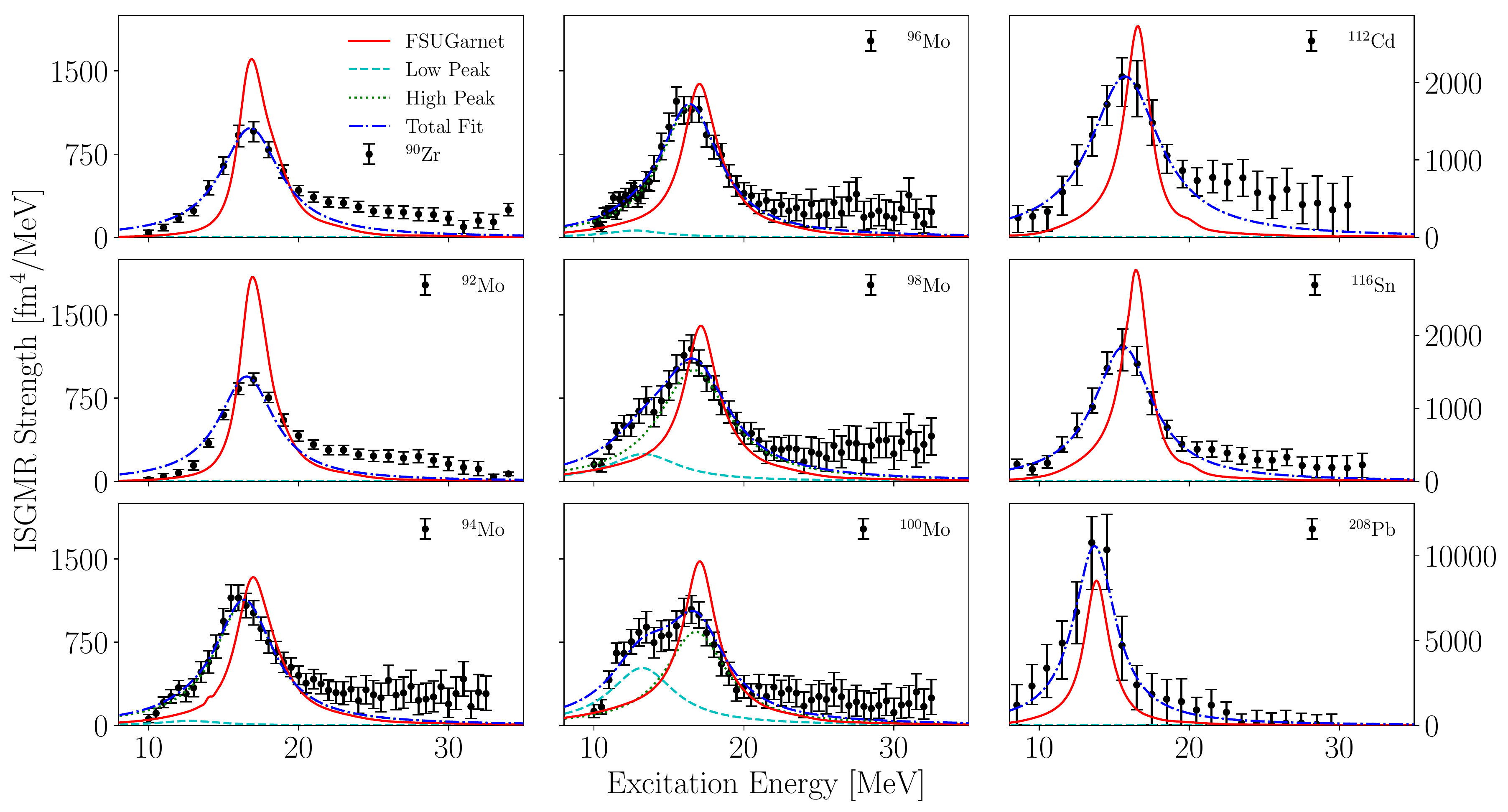}
  \caption{(Color online) ISGMR strength distributions extracted within the MDA framework for nuclei in the present work, as well as for \nuc{90}{Zr} and \nuc{92}{Mo}~\cite{gupta_A90_plb}, \nuc{112}{Cd}~\cite{patel_cd}, \nuc{116}{Sn}~\cite{Li_PRL,Li_PRC}, and \nuc{208}{Pb}~\cite{patel_MEM}. RPA calculations using the FSUGarnet interaction are displayed, as also the one- or two-peak Lorentzian distributions fitted to the experimental data.}
  \label{strengths_with_theory}
\end{figure*}

Inelastic-scattering data were extracted over broad angular ($0^\circ$--$10^\circ$ in the laboratory frame) and excitation-energy ($10$--$32$ MeV) ranges. The horizontal positions were calibrated using high-quality reference \nuc{24}{Mg}($\alpha,\alpha^\prime$) spectra at each angular and magnetic field setting of the spectrometer. The angular distributions were extracted for each 500-keV energy bin using the background-subtracted spectra. Typical forward-angle spectra are presented in Fig. \ref{molly_spec}.

To extract the giant resonance strength distributions from the experimentally measured angular distributions, one needs optical-model parameters to use in performing the Distorted-Wave Born Approximation (DWBA) calculations for various multipoles, which contribute to the measured inelastic scattering spectra. To constrain these optical-model parameters, angular distributions were measured for elastic scattering off \nuc{98}{Mo}. These distributions were then fitted using the nuclear reactions code PTOLEMY. The single-folding optical model employed in the current work is adopted from Ref.~\cite{khoa_satchler_single_folding}, while the target nuclear densities were taken from the empirical distributions presented in Ref.~\cite{fricke_ADNDT}. The predictive power of the optical-model-parameter set can be assessed via comparison of the corresponding DWBA output with the experimentally extracted angular distributions for $0_1^+\rightarrow 2_1^+$ and $0_1^+ \rightarrow 3_1^-$ transitions. Results of these calculations and the optical-model parameters have been provided in Ref.~\cite{KBH_EPJA,KBH_thesis}.

With these optical model parameters, DWBA calculations were carried out for use in a multipole decomposition analysis (MDA) of the spectra. In this procedure, the experimental angular distributions are decomposed into a linear combination of the DWBA calculations of various multipolarities~\cite{Li_PRL,Li_PRC,KBH_EPJA,KBH_thesis,itoh_sm_nucA,itoh_sm_PRC}:

\begin{align}
\frac{\diff^{2} \sigma^\text{exp.}(\theta_\text{cm},E_x)}{\diff \Omega \, \diff E_x} &= \sum_{\lambda} A_\lambda(E_x) \frac{\diff^2 \sigma^\text{DWBA}(\theta_\text{cm},E_x)}{\diff \Omega \, \diff E_x}.
\end{align}

\noindent If the DWBA calculations presume that the full EWSR is exhausted, then the coefficients $A_\lambda(E_x)$ correspond to the distribution of the EWSR over the excitation energy range; the corresponding strength distributions are readily calculated therefrom using the known expressions for the EWSRs~\cite{harakeh_book}. Further details on the implementation of the MDA for these experimental data are given in Refs.~\cite{KBH_EPJA,KBH_thesis}. Representative MDA results for $^{94}$Mo were presented in Fig. 5 of Ref.~\cite{KBH_EPJA}, and the full compilation of all angular distributions and associated decompositions is available in Ref. \cite{KBH_thesis}.

\begin{table*}[t!]
  \centering
  \caption{Fit parameters for the ISGMR response in the molybdenum nuclei along with the integrated \%EWSR values associated with the fitted peaks up to an excitation energy of $35$ MeV. One-peak Lorentzian fit parameters are also provided for the observed ISGMR strength distributions in \nuc{90,92}{Zr}, \nuc{92}{Mo}~\cite{gupta_A90_PRC}, \nuc{112}{Cd}~\cite{patel_cd}, \nuc{116}{Sn}~\cite{Li_PRL,Li_PRC}, and \nuc{208}{Pb}~\cite{patel_MEM}.}
  \vspace*{2mm}
  \begin{tabular}{@{}ccccccccccc@{}}
  \toprule
               &  & \multicolumn{3}{c}{Low Peak}                       &  & \multicolumn{3}{c}{High Peak} & & Total           \\ \cmidrule(lr){3-5} \cmidrule(l){7-9} \cmidrule(l){11-11}
        &  & $E_0$       & $\Gamma$   & $m_1$             &  & $E_0$       & $\Gamma$ & $m_1$   & &  $m_1$ \\

        &  & [MeV]      &  [MeV]  &  [\%]            &  &  [MeV]      & [MeV]  & [\%]  & & [\%]\\

\cmidrule(lr){3-5} \cmidrule(l){7-9} \cmidrule(l){11-11}
  \nuc{94}{Mo} &  & $12.7 \pm 0.5$ & $2.4 \pm 0.4$ & \asymmerror{2}{3}{2} &  & $16.4 \pm 0.2$ & $2.4 \pm 0.4$  & $86 \pm 3$ & & $88 \pm 4$\\
  \nuc{96}{Mo} &  & $12.7 \pm 0.5$  & $2.3 \pm 0.3$ & \asymmerror{4}{3}{4} &  & $16.4 \pm 0.2$ & $2.4 \pm 0.3$ & $89 \pm 3$  & & $93 \pm 4$\\
  \nuc{98}{Mo} &  & $13.3 \pm 0.5$ & $2.8 \pm 0.5$ & $16 \pm 4$ &  & $16.7 \pm 0.4$ & $2.8 \pm 0.4$  & $85 \pm 4$ & & $102 \pm 6$\\
  \nuc{100}{Mo} &  & $13.2 \pm 0.4$ & $2.6 \pm 0.6$ & $32 \pm 4$ &  & $16.8 \pm 0.4$ & $2.5 \pm 0.5$  & $60 \pm 3$ & & $93 \pm 6$\\
  \cmidrule(l){1-11}
  \nuc{90}{Zr} &  & --                & --               & --                    &  & $16.8 \pm 0.2$ & $2.4 \pm 0.4$ & $84 \pm 2$ & & $84\pm2$ \\
\nuc{92}{Zr} &  & --               & --               & --                    &  & $16.4 \pm 0.1$ & $2.2 \pm 0.3$ & $91 \pm 2$ & & $91 \pm 2$ \\
\nuc{92}{Mo} &  & --               & --               & --                    &  & $16.5 \pm 0.1$ & $2.3 \pm 0.1$ & $73 \pm 2$ & & $73 \pm 2$\\
  \nuc{112}{Cd} &  & -- & -- & -- &  & $15.8 \pm 0.2$ & $2.8 \pm 0.3$  & $137 \pm 8$ & & $137 \pm 8$\\
  \nuc{116}{Sn} &  & -- & -- & -- &  & $15.6 \pm 0.1$ & $2.4 \pm 0.2$  & $90 \pm 3 $ & & $90 \pm 3$\\
  \nuc{208}{Pb} &  & -- & -- & -- &  & $13.7 \pm 0.1$ & $1.7 \pm 0.1$  & $140 \pm 3$ & & $140 \pm 4$\\
  \bottomrule
  \end{tabular}

\label{fit_parameters}
\end{table*}

The extracted ISGMR strength distributions for \nuc{94,96,98,100}{Mo} are given in Fig. \ref{strengths_with_theory}, in addition to ISGMR strength extracted in \nuc{90}{Zr}~\cite{gupta_A90_PRC}, \nuc{92}{Mo}~\cite{gupta_A90_PRC}, \nuc{112}{Cd}~\cite{patel_cd}, \nuc{116}{Sn}~\cite{Li_PRL,Li_PRC}, and \nuc{208}{Pb}~\cite{patel_MEM}; all of these have been measured at RCNP, using identical experimental and analytical methodologies as presented here. Also shown are results from RPA calculations using the relativistic FSUGarnet interaction~\cite{fsugarnet_commissioned} for each nucleus in question. FSUGarnet belongs to a class of covariant energy density functionals that are informed by the properties of both finite nuclei and neutron stars. In particular, the model parameters of FSUGarnet were calibrated to the binding energies and charge radii of magic and semi-magic nuclei, ISGMR centroid energies in \nuc{90}{Zr} and \nuc{208}{Pb}, and current limits on the maximum mass of neutron stars. As such, FSUGarnet aims to describe within a unified framework nuclear phenomena that happen at length scales that vary by more than 18 orders of magnitude. It should be noted that FSUGarnet has $K_\infty\!=\!229.6\!\pm\!2.5$ MeV.

It is qualitatively evident from the direct comparison of the presented experimental and theoretical strength distributions that the FSUGarnet interaction is able to reproduce quite precisely the excitation energy of the ISGMR in \nuc{208}{Pb}. In contrast, the analogous calculations for \nuc{112}{Cd} and \nuc{116}{Sn} overestimate the ISGMR energies in a manner that is consistent with the premise that these isotopes are soft, as has been previously discussed. Furthermore, the ISGMR strengths in the molybdenum nuclei --- in particular, \nuc{94,96}{Mo} --- appear to be manifesting exactly the same behavior inasmuch that the RPA calculations tend to peak at the high-energy shoulders of the experimentally extracted strength distributions. Indeed, one finds that the difference between the peak positions as predicted by using the FSUGarnet interaction relative to the experimental values are nearly identical in the case of \nuc{94,96}{Mo}, \nuc{112}{Cd}, and \nuc{116}{Sn}. This provides clear evidence that one cannot simultaneously reproduce the ISGMR in the molybdenum nuclei and \nuc{208}{Pb}, which is precisely the same predicament that had plagued the description of the ISGMR in the tin and cadmium chains~\cite{piek_physG,Piekarewicz_centelles}.

To quantify this effect further, Table \ref{fit_parameters} lists the fit parameters, which model the ISGMR strength distributions of the molybdenum nuclei using a Lorentzian line-shape:
\begin{align}
    S(E_x,S_0,E_{0},\Gamma) & = \frac{S_0 \Gamma}{\left(E_x - E_0\right)^2 + \Gamma^2}.
    \label{lorentz}
\end{align}

As is suggested by the experimental strengths presented in Fig. \ref{strengths_with_theory}, the ISGMR response of \nuc{98,100}{Mo} is best modeled by a two-peak shape that is conjectured to arise from the effects of deformation. This low-energy peak arises due to coupling between the $K=0$ component of the ISGQR and the main ISGMR \cite{garg_sm_PRL,zawischa_deformation,kvasil_deformation_analysis}; the strength of this coupling is generally understood to increase with larger ground-state deformation. This structure effect is well-documented in various regions of the nuclear chart~\cite{itoh_sm_PRC,itoh_sm_nucA,garg_sm_PRL,zawischa_deformation,kvasil_deformation_analysis,gupta_24Mg_plb,peach_28Si_prc}, and it is unrelated to the question of softness that is the focus of the present work.

The RPA calculations with the FSUGarnet interaction are spherical in nature and do not allow for the aforementioned deformation degrees of freedom. In the experimental analysis of the strengths, however, the possibility of a low-energy peak arising due to deformation was included in the extraction of the fit parameters presented in Table \ref{fit_parameters}.\footnote{N.B. the development of deformation in \nuc{98,100}{Mo} causes a modest decrease in the experimentally extracted ISGMR centroid energies, which remains unaccounted for in the presented spherical RPA calculations.} In the cases of \nuc{94,96}{Mo}, it was determined that the amounts of the EWSR exhausted in the low-energy peak due to deformation were consistent with $0\%$ and therefore  the deformation effects on the ``main'' ISGMR
peaks are negligible as far as comparisons with other spherical nuclei are concerned~\cite{KBH_EPJA}.

\begin{figure}[h!]
  \includegraphics[width=\linewidth]{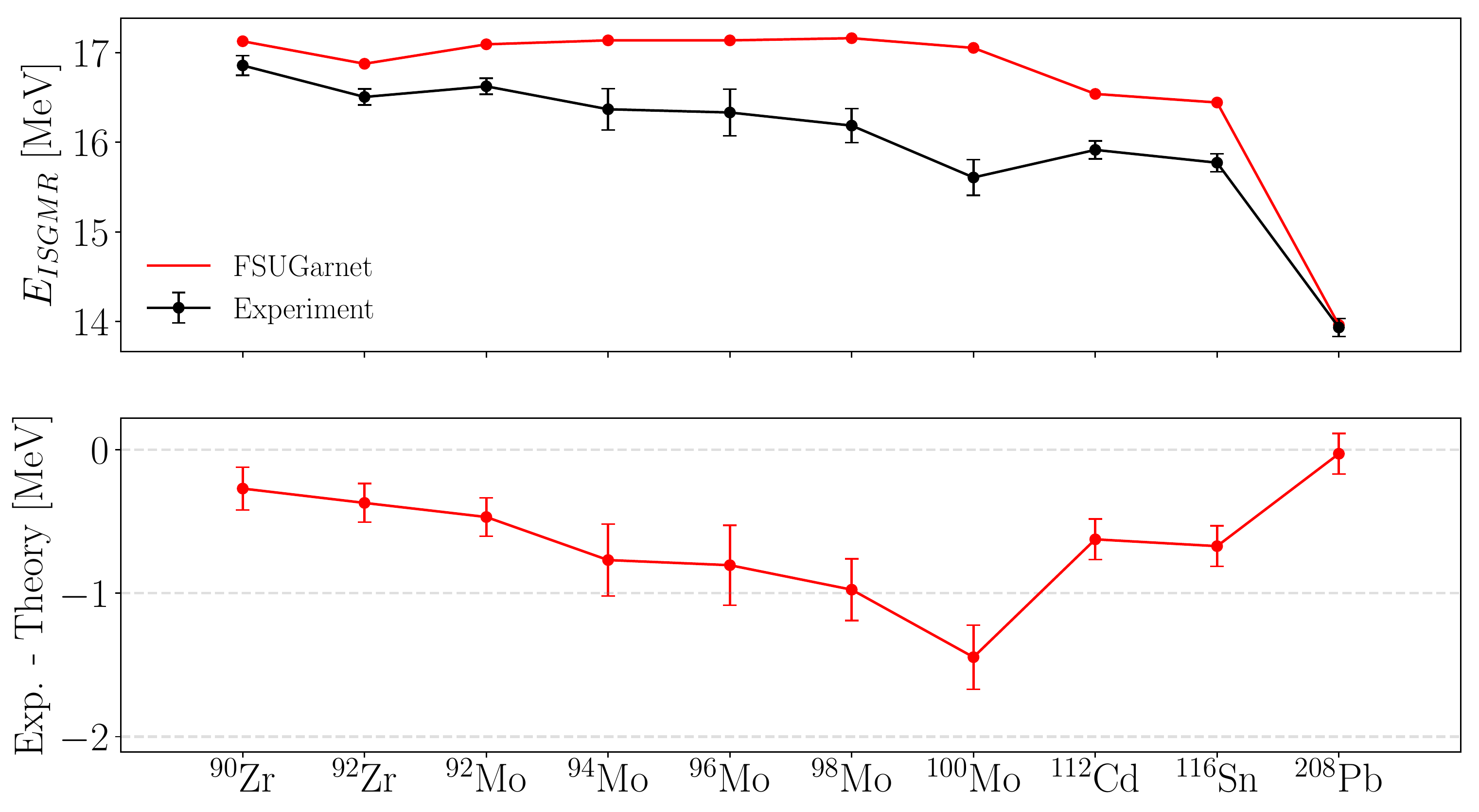}
  \caption{(Color online) Top: experimentally measured and theoretically calculated centroid energies ($m_1/m_0$) for the molybdenum isotopes studied here, relative to those of \nuc{90}{Zr},~\cite{gupta_A90_PRC}, \nuc{92}{Zr},~\cite{gupta_A90_PRC}, \nuc{92}{Mo}~\cite{gupta_A90_PRC}, \nuc{112}{Cd}~\cite{patel_cd}, \nuc{116}{Sn}~\cite{Li_PRL,Li_PRC}, and \nuc{208}{Pb}~\cite{patel_MEM}. Bottom: relative difference between the centroid energies predicted by using the FSUGarnet interaction and those extracted from experiment for each of the nuclei (theoretical uncertainties are included in the presented error bars and are of the order $\sim 100$ keV)~\cite{Chen:2014sca}.}
  \label{moments_comparison}
\end{figure}

The conventional moment ratios ($\sqrt{m_1/m_{-1}}$, $m_1/m_0$, and $\sqrt{m_3/m_1}$), calculated over the excitation-energy range $0$ -- $35$ MeV from the fitted experimental strength distributions, are presented in
Table \ref{molly_moment_ratios_total_EWSRS}. A comparison of these moment ratios with those extracted in the same manner from the FSUGarnet calculations is made in Fig. \ref{moments_comparison}.

\begin{table}[h!]
\centering
\caption{Experimentally extracted moment ratios for \nuc{94-100}{Mo} calculated between $0$\,--\,$35$
MeV from the fit distributions of Table \ref{fit_parameters}. Also presented are the corresponding moment ratios from the ISGMR data of \nuc{90,92}{Zr}, \nuc{92}{Mo}~\cite{gupta_A90_PRC}, \nuc{112}{Cd}~\cite{patel_cd}, \nuc{116}{Sn}~\cite{Li_PRL,Li_PRC}, and \nuc{208}{Pb}~\cite{patel_MEM} which have been re-evaluated using the distributions of Table \ref{fit_parameters} over this energy range.}
\vspace*{2mm}
\begin{tabular}{@{}cccc@{}}
\toprule
Nucleus      & $\sqrt{m_{1}/m_{-1}}$ & $m_1/m_0$   & $\sqrt{m_{3}/m_{1}}$   \\
     & [MeV] & [MeV]  &  [MeV]  \\ \midrule
\nuc{94}{Mo} & $15.2 \pm 0.3$            & $16.4 \pm 0.2$ & $18.5 \pm 0.5$           \\
\nuc{96}{Mo} & $15.2 \pm 0.3$            & $16.3 \pm 0.2$ & $18.4 \pm 0.4$            \\
\nuc{98}{Mo} & $14.8 \pm 0.3$            & $16.2 \pm 0.2$ & $18.7 \pm 0.7$            \\
\nuc{100}{Mo} & $14.3 \pm 0.4$            & $15.6 \pm 0.2$ & $18.1 \pm 0.7$            \\
\midrule
\nuc{90}{Zr} & $15.7\pm0.1$ & $16.9 \pm 0.1$ & $18.9 \pm 0.2$ \\
\nuc{92}{Zr} & $15.2\pm0.1$ & $16.5 \pm 0.1$ & $18.7 \pm 0.1$ \\
\nuc{92}{Mo} & $15.5\pm0.1$ & $16.6 \pm 0.1$ & $18.6 \pm 0.1$ \\
\nuc{112}{Cd} & $14.6\pm0.1$ & $15.9 \pm 0.1$ & $18.4 \pm 0.1$ \\
\nuc{116}{Sn} & $14.6\pm0.1$ & $15.8 \pm 0.1$ & $17.9 \pm 0.1$ \\
\nuc{208}{Pb} & $13.1\pm0.1$ & $13.9 \pm 0.1$ & $15.8 \pm 0.2$ \\ \bottomrule
\end{tabular}
\label{molly_moment_ratios_total_EWSRS}
\end{table}

It is important to note here that while the predicted ISGMR centroid energy of \nuc{208}{Pb} is consistent within experimental uncertainties, the amounts by which the centroid energies of \nuc{112}{Cd} and \nuc{116}{Sn} are overestimated by using the FSUGarnet interaction are consistent with the corresponding overestimation of the ISGMR in \nuc{94,96}{Mo}. One can, therefore, conclude that the molybdenum nuclei are also ``soft'' in precisely the same way as has been documented in the tin and cadmium nuclei and this ``softness'' begins as early as moving just two nucleons away from the
``doubly-magic" nucleus \nuc{90}{Zr}.

Results of a detailed theoretical analysis of the ISGMR and ISGQR strength distributions in the molybdenum isotopes investigated in this work within the quasiparticle-RPA framework, taking into account the pairing correlations and the effects of axial deformation, have recently become available \cite{colo_nesterenko_private}. The authors discuss at length the effects of four Skyrme interactions (SkM*, SLy6, SVbas, and SkP$^{\delta}$) on the modeling of the ISGMR and the nuclear incompressibility within the well-accepted microscopic methodology of Ref. \cite{blaizot}. The analysis concludes that the deformation-induced coupling between the ISGMR and ISGQR plays a critical role in reproducing the observed ISGMR strengths. Their results are, nevertheless, not inconsistent with the onset of ``softness'' in the molybdenum isotopes discussed in the present work: they find that the best reproduction of the ISGMR data in the molybdenum nuclei comes from the SKP$^{\delta}$ interaction, which has $K_\infty = 202$ MeV, a value that is significantly lower than the currently acceptable value for this quantity ($K_\infty = 240 \pm 20$ MeV).

In summary, we have measured the ISGMR strength distributions in the even-A molybdenum isotopes ($A=94$--$100$). The ISGMR responses of these nuclei appear to  suggest a value for
$K_\infty$ significantly lower than the currently accepted value of $K_\infty = 240 \pm 20$ MeV, similar to what has previously been documented in the isotopic chains of tin and cadmium. The softness appears to gradually increase with the addition of nucleons to \nuc{90}{Zr}, before manifesting fully in the case of \nuc{94}{Mo}. Calculations with a modern relativistic interaction, FSUGarnet, which reproduces the ISGMR energy of \nuc{208}{Pb} well, are unable to reproduce the centroids of the ISGMR strengths for these molybdenum nuclei, clearly pointing to their ``softness''.  The question: ``\emph{Why are the tin isotopes soft?}'' remains unresolved and makes further exploration of this phenomenon most imperative.

KBH acknowledges the support of the Arthur J. Schmitt Foundation, as well as the Liu Institute for Asia and Asian Studies, University of Notre Dame. This work has been supported in part by the National Science Foundation (Grant No. PHY-1713857) and the U.S. Department of Energy Office of Nuclear Physics under Award Number DE-FG02-92ER40750.

\bibliography{moly_plb}

\end{document}